\documentclass[sn-mathphys-num]{sn-jnl}

\usepackage[abs]{overpic}
\usepackage{graphicx}%
\usepackage{multirow}%
\usepackage{amsmath,amssymb,amsfonts}%
\usepackage{amsthm}%
\usepackage{mathrsfs}%
\usepackage[title]{appendix}%
\usepackage{xcolor}%
\usepackage{textcomp}%
\usepackage{manyfoot}%
\usepackage{booktabs}%
\usepackage{algorithm}%
\usepackage{algorithmicx}%
\usepackage{algpseudocode}%
\usepackage{listings}%
\usepackage{scalerel}
\usepackage{tikz}
\usetikzlibrary{svg.path}
\usepackage[abs]{overpic}
\usepackage{hyperref}
\usepackage{graphicx}%
\usepackage{multirow}%
\usepackage{amsmath,amssymb,amsfonts}%
\usepackage{amsthm}%
\usepackage{mathrsfs}%
\usepackage[title]{appendix}%
\usepackage{xcolor}%
\usepackage{textcomp}%
\usepackage{manyfoot}%
\usepackage{booktabs}%
\usepackage{algorithm}%
\usepackage{algorithmicx}%
\usepackage{algpseudocode}%
\usepackage{listings}%
\usepackage{scalerel}
\usepackage{tikz}
\usetikzlibrary{svg.path}
\usepackage{natbib}
\usepackage{algorithm}
\usepackage{algpseudocode}
\usepackage{bm}

\definecolor{orcidlogocol}{HTML}{A6CE39}
\tikzset{
  orcidlogo/.pic={
    \fill[orcidlogocol] svg{M256,128c0,70.7-57.3,128-128,128C57.3,256,0,198.7,0,128C0,57.3,57.3,0,128,0C198.7,0,256,57.3,256,128z};
    \fill[white] svg{M86.3,186.2H70.9V79.1h15.4v48.4V186.2z}
                 svg{M108.9,79.1h41.6c39.6,0,57,28.3,57,53.6c0,27.5-21.5,53.6-56.8,53.6h-41.8V79.1z M124.3,172.4h24.5c34.9,0,42.9-26.5,42.9-39.7c0-21.5-13.7-39.7-43.7-39.7h-23.7V172.4z}
                 svg{M88.7,56.8c0,5.5-4.5,10.1-10.1,10.1c-5.6,0-10.1-4.6-10.1-10.1c0-5.6,4.5-10.1,10.1-10.1C84.2,46.7,88.7,51.3,88.7,56.8z};
  }
}

\newcommand\orcidicon[1]{\href{https://orcid.org/#1}{\mbox{\scalerel*{
\begin{tikzpicture}[yscale=-1,transform shape]
\pic{orcidlogo};
\end{tikzpicture}
}{|}}}}

\usepackage{hyperref}

\theoremstyle{thmstyleone}%
\theoremstyle{thmstyletwo}%

\theoremstyle{thmstylethree}%

\begin{document}

\title[Article Title]{A meshless method to compute the proper orthogonal decomposition and its variants from scattered data}


\author*[1]{\fnm{Iacopo} \sur{Tirelli}}\email{iacopo.tirelli@uc3m.es \orcidicon{0000-0001-7623-1161}}

\author[2]{\fnm{Miguel Alfonso} \sur{Mendez}}\email{mendez@vki.ac.be \orcidicon{0000-0002-1115-2187}}

\author[1]{\fnm{Andrea} \sur{Ianiro}}\email{aianiro@ing.uc3m.es \orcidicon{0000-0001-7342-4814}}

\author[1]{\fnm{Stefano} \sur{Discetti}}\email{sdiscett@ing.uc3m.es \orcidicon{0000-0001-9025-1505}}

\affil*[1]{\orgdiv{Department of Aerospace Engineering}, \orgname{Universidad Carlos III de Madrid}, \orgaddress{\street{Avda. Universidad 30}, \city{Leganés}, \postcode{28911}, \state{Madrid}, \country{Spain}}}

\affil[2]{\orgdiv{Environmental and Applied Fluid Dynamics}, \orgname{von Karman Institute for Fluid Dynamics}, \orgaddress{\street{Waterloosesteenweg 72}, \city{Sint-Genesius-Rode}, \postcode{1640}, \state{Bruxelles}, \country{Belgium}}}





\abstract{
	Complex phenomena can be better understood when broken down into a limited number of simpler ``components''. Linear statistical methods such as the principal component analysis and its variants are widely used across various fields of applied science to identify and rank these components based on the variance they represent in the data. These methods can be seen as factorisations of the matrix collecting all the data, assuming it consists of time series sampled from fixed points in space. However, when data sampling locations vary over time, as with mobile monitoring stations in meteorology and oceanography or with particle tracking velocimetry in experimental fluid dynamics, advanced interpolation techniques are required to project the data onto a fixed grid before the factorisation. This interpolation is often expensive and inaccurate. This work proposes a method to decompose scattered data without interpolating. The approach employs physics-constrained radial basis function regression to compute inner products in space and time. The method provides an analytical and mesh-independent decomposition in space and time, demonstrating higher accuracy. Our approach allows distilling the most relevant ``components'' even for measurements whose natural output is a distribution of data scattered in space and time, maintaining high accuracy and mesh independence.
}


\maketitle

\section{Introduction}
\label{Sec:Intro}

Methods from dimensionality reduction and pattern identification have become ubiquitous in many applied sciences, allowing streamlining data analysis, simplifying complex datasets, highlighting essential features and uncovering critical patterns and relationships \cite{Holmes_Lumley_Berkooz_1996, Martinson_2018}. 

\textcolor{black}{Although the rise of machine learning has introduced new approaches for nonlinear reduced-order modeling of PDEs using artificial neural networks \cite{Hesthaven2018, Muecke2021, Fresca2022, brivio2024ptpi}, linear dimensionality reduction methods continue to serve as the foundation of equation-free reduction techniques for data processing and compression \cite{mendez_2023_chap8}, face recognition \citep{kirby1990application}, gap filling in experimental data \cite{saini2016development}, adaptive least square problems \citep{yao2017empirical}, fault diagnostics \citep{shen2022informative}, optimal sensor placement \citep{castillo2020data}.}

The simplest and fundamental approach to linear dimensionality reduction is the Principal Component Analysis (PCA), originally formulated in multivariate statistics \cite{pearson1901principal,hotelling1936simplified} and known as Karhunen-Loève transform (KL) in mathematics \cite{schmidt1907theorie}, Empirical Orthogonal Functions (EOF) in meteorology and climatology \citep{obukhov1947statistically,lorenz1956empirical,kutzbach1967empirical} and Proper Orthogonal Decomposition (POD) in fluid mechanics \cite{lumley1967structure,sirovich1987turbulence,sirovich1991analysis}. PCA simplifies large datasets by decomposing them into a set of linearly uncorrelated variables known as principal components, ordered by the amount of variance they capture from the data. Its simplicity, versatility and effectiveness have established the PCA as one of the crucial first steps in many data analysis workflows, and many extensions and variants have been proposed \cite{Ghojogh2019,Ghojogh2019a,Schoelkopf1997,Jolliffe2016,Mendez2023_MST}.

Beyond multivariate statistical analysis, principal components are often used to reveal spatial correlation of time series. In fluid mechanics, for example, the extension of PCA onto POD aimed at using such dimensionality reduction methods to identify coherent structures in turbulent flows \citep{lumley1967structure,Holmes_Lumley_Berkooz_1996,jimenez_2023}. In computational physics, the Galerkin projection of a partial differential equations onto these modes is the cornerstone to reduce the computational cost of large simulations \cite{Benner2015}, to build a reduced order model that enables model-based control \cite{Ahmed2021,Bernd2011a} and to derive more efficient Large Eddy Simulation formulations \cite{Girfoglio2021}.
In climatology, the EOF \cite{hannachi2007empirical,Monahan2009,Martinson_2018} has been developed to identify and analyse dominant patterns of variability in climate data and to reduce the dimensionality of large climate datasets. In all these applications, the leading patterns are identified as eigenvectors of the correlation matrices, computed in space or in time. However, the computation of these matrices implies the definition of a grid, to approximate the integral in the continuous formulation of the problems \cite{Shen1998,Errors_EOF}. A major research effort in both fluid mechanics and climatology is the problem of handling missing data from these grids \cite{Everson1995,EOF_Sparse,Willcox2004,Tseringxiao2019}.
Common approaches to handle these cases are the Gappy POD \cite{Everson1995}, Data Interpolating EOF  \citep{Alvera‐Azcarate2007}, or probabilistic PCA
\citep{Tipping1999}. An extensive overview of gap-filling methods is provided in Refs.~\cite{goodin1979comparison,Miro2017}.

When the level of missing data is extreme, the concept of a grid becomes irrelevant and data could be treated as randomly scattered. This is the natural condition in image velocimetry measurements carried out with particle tracking methods \cite{Schanz2016,Tan2020,Schroeder2023}, in the measurement of wind or concentration fields from limited stations \cite{goodin1979comparison,Miro2017} or in the interpolation of geological data from limited locations \cite{wang2017interpolation}.

This article proposes a grid-less approach to the computation of POD/EOF. The idea consists of computing correlation matrices of scattered data using constrained Radial Basis Functions (RBFs) as introduced in Ref.~\cite{sperotto2022meshless}, together with a Gauss-Legendre quadrature for the integration. We show that the approach provides higher accuracy than the usual interpolation-based approach and provides an analytical (mesh-independent) representation of the spatial structures. \textcolor{black}{The POD computation based on the smoother or interpolator operator of the data makes the proposed approach closely related to the Functional PCA \citep[FPCA,][]{ramsay2005principal,
		Ramsay1997, Wang2016, Hall2006}, which extends the PCA to functional data, where each observation is a continuous function. The key difference, however, is that the proposed approach preserves the finite nature of the covariance matrices, keeping the algorithmic structure of the dual PCA \cite{Tripathy2021,Ghojogh2022} and snapshot POD \cite{sirovich1987turbulence,sirovich1991analysis}. This similarity is further discussed in \S~\ref{subsec:parametric_functional}.}

The theoretical background of the proposed methodology is detailed in  \S~\ref{sub:algorithm}, while the validation is presented in \S~\ref{Sec.validation} on two different synthetic datasets: an analytical sinusoidal test case (\S~\ref{sub.sine}) and the wake of a fluidic pinball (\S~\ref{Sec.pinball}). An experimental validation is carried out in \S~\ref{sub.meteo} using the precipitation data collected from satellite microwave observations.

\section{Methodology}\label{Sec.methodology}

The proposed methodology follows the classic snapshot-based POD of Sirovich \cite{sirovich1991analysis}, but replaces all inner products with an RBF formulation that does not require a grid and thus does not require interpolation. The data can be scattered in space and time, and here we assume it to be vector valued.
The final result is a linear decomposition of the data in the following form:

\begin{equation}
	\bm{u}(\bm{x}, {t}) = \sum^{r_c}_{r=1}\sigma_r \bm{\phi}_r(\bm{x})\psi_r({t})\,,
	\label{POD}
\end{equation} 
where \textcolor{black}{$\bm{x}\in\mathbb{R}^d$ (with $d$ the dimensionality of the problem) and $t$ are respectively the spatial and the temporal coordinate,} $r_c$ is the rank, $\sigma_r$ are scalars defining the amplitude of each contribution, $\bm{\phi}_r(\bm{x})$ and  $\psi_r({t})$ are the spatial and the temporal structures of the POD modes.  We here assume that $\bm{u}(\bm{x},t)\in\mathbb{R}^{d}$ is a vector field take $\bm{\phi}_r(\bm{x})\in\mathbb{R}^{d}$ to be of the same dimension, that is the temporal structures is common to all its components.

\subsection{The meshless POD algorithm} \label{sub:algorithm}

\begin{figure}[t]
	\centering
	\begin{overpic}[scale =0.8, unit= 1mm]
		{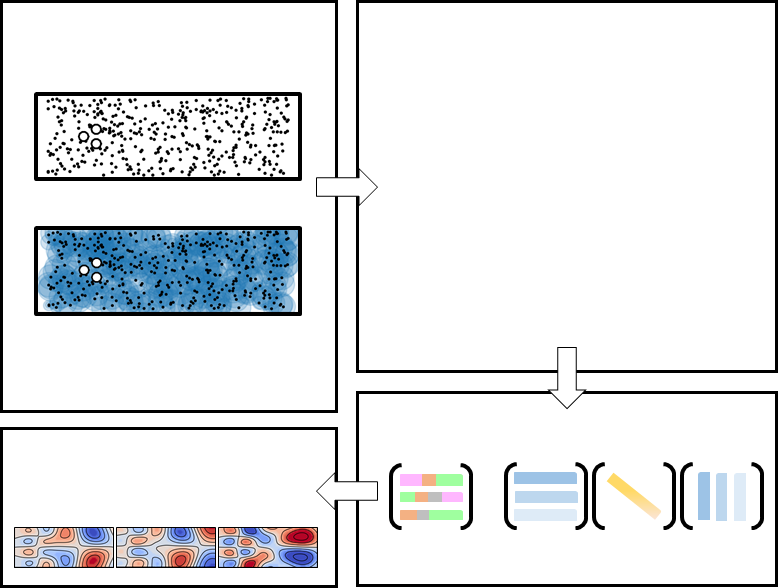}
		\put(0,90){\parbox{58mm}{\centering \textnormal{\textbf{1.  Analytical snapshot representation with RBF
		}}}}
		\put(29, 67.5){\vector(0, -1){5}}  
		\put(33, 63.5){{for each $\mathbf{t}_i$}} 
		\put(0,36){\parbox{58mm}{\centering \textnormal{$\tilde{u}(\bm{x}, \mathbf{t}_i) = \sum_{q=1}^{N_b^{(i)}} w_q(\mathbf{t}_i) \gamma_q(\bm{x};\mathbf{X}_q^{(i)})$
		}}}
		\put(60,90){\parbox{70mm}{\centering \textnormal{\textbf{2.  Temporal correlation matrix
		}}}}
		
		\put(60,75){\parbox{70mm}{\centering \textnormal{\scalebox{0.8}{$\mathbf{K} = \begin{bmatrix}
							\langle \tilde{\bm{u}}_1, \tilde{\bm{u}}_1 \rangle_s & \langle \tilde{\bm{u}}_1, \tilde{\bm{u}}_2 \rangle_s & \dots & \langle \tilde{\bm{u}}_1, \tilde{\bm{u}}_{N_t} \rangle_s \\
							\langle \tilde{\bm{u}}_2, \tilde{\bm{u}}_1 \rangle_s & \langle \tilde{\bm{u}}_2, \tilde{\bm{u}}_2 \rangle_s & \dots & \langle \tilde{\bm{u}}_2, \tilde{\bm{u}}_{N_t} \rangle_s \\
							\vdots & \vdots & \ddots & \vdots \\
							\langle \tilde{\bm{u}}_{N_t}, \tilde{\bm{u}}_1 \rangle_s & \langle \tilde{\bm{u}}_{N_t}, \tilde{\bm{u}}_2 \rangle_s & \dots & \langle \tilde{\bm{u}}_{N_t}, \tilde{\bm{u}}_{N_t} \rangle_s \end{bmatrix}$}}}}
		\put(60, 60){\parbox{70mm}{\centering \textnormal{$K_{ij} = \frac{1}{|\Omega|} \int_\Omega \bm{u}^T(\bm{x}', \mathbf{t}_i) \bm{u}(\bm{x}', \mathbf{t}_j) \, d\bm{x}'$
		}}}
		\put(95, 58){\vector(0, -1){5}}  %
		\put(60, 47){\parbox{70mm}{\centering \textnormal{Gauss-Legendre quadrature\\ \vspace{0.07cm} $K_{ij} = \frac{1}{2^d} \sum_{k=1}^{n} \bm{w}_{L_k}{\bm{\tilde{u}}}(\bm{X}_{L_k},{t_i}) {\bm{\tilde{u}}}(\mathbf{X}_{L_k},{t_j}) $
		}}}
		\put(60,27){\parbox{70mm}{\centering \textnormal{\textbf{3.  Computing temporal structures
		}}}}
		\put(68,5){\parbox{10mm}{\centering {{$\mathbf{{K}}$}}}} 
		\put(80,5){\parbox{5mm}{\centering {{=}}}}
		\put(88,5){\parbox{10mm}{\centering {{$\boldsymbol{{\Psi}}$}}}} 
		\put(103,5){\parbox{10mm}{\centering {{$\boldsymbol{\Sigma}^2$}}}}
		\put(119,5){\parbox{10mm}{\centering {{$\boldsymbol{\Psi}^T$}}}} 
		\put(0,21){\parbox{58mm}{\centering \textnormal{\textbf{4.  Compute spatial structures
		}}}}
		\put(0,14){\parbox{58mm}{\centering \textnormal{{$\bm{\phi}_r(\bm{x}) = \frac{1}{\sigma_r}\langle \bm{\tilde{u}}(\bm{x}, {t}), \psi_r({t}) \rangle$
		}}}}
		
	\end{overpic}
	\caption{\textcolor{black}{Flowchart of the proposed algorithm: in the first step, analytical approximation of the snapshots are obtained via RBF from the scattered data for each time instant. Subsequently, these representations are employed to compute the temporal correlation matrix $\mathbf{K}$ in its continuous form, with integrals computed through the Gauss-Legendre quadrature. Once the entire matrix is filled, an eigendecomposition provides the mesh-free temporal modes collected in the matrix $\boldsymbol{\Psi}$ and the eigenvalues matrix $\boldsymbol{\Sigma}$. In the last step, the spatial modes $\boldsymbol{\phi}$ are obtained projecting the approximation of the field onto the temporal structures.} }
	\label{Fig:flowchart}
\end{figure}

\textcolor{black}{A graphical summary of the methodology is depicted in Fig.\ref{Fig:flowchart}, while more details are provided in the Algorithm~\ref{pseudo}.} The methodology is made of four steps:

\vspace{0.1cm}
\textit{\textbf{Step 1: Analytical snapshot representation with RBF}}

Starting from scattered data available for the different time realizations, we are interested in an analytical approximation of the field for each time instant as a linear combination of a set of basis functions. Let $\mathbf{t} = [t_1, \dots, t_{N_t}]$ denote the vector that collects all time steps at which a set of $\mathbf{X}^{(k)} \in \mathbb{R}^{N_p(k) \times d }$ randomly scattered points is available. Note that the time steps are not necessarily uniform, and both the number and the locations of samples vary across snapshots. We use the notation $\mathbf{X}^{(k)}_q=\mathbf{X}^{(k)}[q,:]\in\mathbb{R}^{d}$ to refer to a specific sample in the set of points available at the $k$-th snapshot. We seek an approximation of each snapshot as:

\begin{equation}
	\bm{\tilde{u}}(\bm{x}, \mathbf{t}_k) = \sum_{q=1}^{N_b^{(k)}} \bm{w}_q(\mathbf{t}_k) {\gamma}_q(\bm{x};\mathbf{X}_q^{(k)}),
	\label{func}
\end{equation}

\noindent where $N_b^{(k)}$ is the number of basis functions used in the approximation of the snapshot $k$, $\boldsymbol{\gamma}_q$ is the $q$-{th} regression basis, and $\bm{w}_q\in\mathbb{R}^{3}$ is the weight vector that gives the contribution of the $q$-th basis to each of the components of the velocity field. With no loss of generality, the basis functions employed in this work are thin-plate spline RBF \textcolor{black}{\cite{buhmann2000radial}, defined as follow:}

\begin{equation} 
	\textcolor{black}{\gamma_q(\bm{x};\mathbf{X}_q^{(k)})=\gamma_q({r}(\bm{x};\mathbf{X}_q^{(k)})) = {r}^2\log({r})\,,}
	\label{eq.thinplate}
\end{equation} with ${r} = \|\bm{x} - \mathbf{X}_q^{(k)}\|$ the radial distance from the  collocation point. Note that the available points $\mathbf{X}^{(k)}$ are used here as collocation points; however, this need not be the case in general, and different collocation points could be used. Since these bases have no shape factor, the current approach has no hyper-parameters to tune, needs no user input and only requires extra memory storage for the weights. Future work will explore more advanced formulations and the use of physics constraints as in Ref.~\cite{sperotto2022meshless}.

\vspace{0.1cm}
\textit{\textbf{Step 2: Temporal correlation matrix}}

The temporal correlation matrix $\mathbf{K}\in\mathbb{R}^{N_t \times N_t}$ is defined as the matrix whose elements are the inner products in space, denoted as $\langle \cdot, \cdot \rangle_s$, between all the snapshots:
\begin{equation}
	\mathbf{K} = \begin{bmatrix}
		\langle \bm{u}_1, \bm{u}_1 \rangle_s & \langle \bm{u}_1, \bm{u}_2 \rangle_s & \dots & \langle \bm{u}_1, \bm{u}_{N_t} \rangle_s \\
		\langle \bm{u}_2, \bm{u}_1 \rangle_s & \langle \bm{u}_2, \bm{u}_2 \rangle_s & \dots & \langle \bm{u}_2, \bm{u}_{N_t} \rangle_s \\
		\vdots & \vdots & \ddots & \vdots \\
		\langle \bm{u}_{N_t}, \bm{u}_1 \rangle_s & \langle \bm{u}_{N_t}, \bm{u}_2 \rangle_s & \dots & \langle \bm{u}_{N_t}, \bm{u}_{N_t} \rangle_s \\
	\end{bmatrix},
	\label{K}
\end{equation}


where the short-hand notation $\bm{u}_k$ here denotes the k-th snapshot, which in general could be a continuous field or a vector that collects data.

In the original formulation  of the POD \citep{lumley1967structure}, the inner product in space or time was defined in terms of correlation of square-integrable real-valued vector fields, that is:

\begin{equation}
	\label{Key}
	K_{ij} = \frac{1}{|\Omega|} \int_\Omega \bm{u}^T(\bm{x}', \mathbf{t}_i) \bm{u}(\bm{x}', \mathbf{t}_j) \, d\bm{x}'\,,
\end{equation}

\noindent with \textcolor{black}{$|\Omega|$} the area (in 2D) or volume (in 3D) of the spatial domain considered. If the spatial domain is partitioned in uniform $N_s$ elements of area (or volume) $\Delta \Omega$, indexed by $k\in[0,N_s-1]$ the most natural approximation of Eq.~\eqref{Key} reads

\begin{equation}
	\label{Key_2}
	K_{ij} \approx \frac{1}{N_s \Delta \Omega} \sum_{k} \bm{u}^T(\mathbf{X}_k, \mathbf{t}_i) \, \bm{u}(\mathbf{X}_k, \mathbf{t}_j) \, \Delta \Omega
	= \frac{1}{N_s} \, \mathbf{u}_j^T \mathbf{u}_i \,,
\end{equation}

\noindent where $\mathbf{u}_i,\mathbf{u}_j\in\mathbb{R}^{d\,N_s}$ are the vectors that collect data sampled in the time steps $\mathbf{t}_i$ and $\mathbf{t}_j$ and corresponding to each of the $N_s$ portions of the spatial domain and $\mathbf{X}_k$ is the k-th entry of the fixed available grid. In this setting, the correlation matrix can be calculated using simple matrix multiplication as $\mathbf{K}=\mathbf{U}^T\mathbf{U}$, with $\mathbf{U}\in\mathbb{R}^{d N_s\times N_t}$ collecting all the data with the velocity components stacked along the columns. \textcolor{black}{It is important to note that obtaining a \textit{centered} correlation matrix requires subtracting the mean from the data set. This approach was used by the authors following the methodology outlined in Ref.~\cite{tirelli2023simple}.}


Eq.~\eqref{Key_2} is an approximation of Eq.~\eqref{Key} using the midpoint rule \citep{Numerics}, that is, assuming a piecewise-constant approximation of the velocity field. \textcolor{black}{This is true only for a uniformly sampled grid (that is, the $\Delta \Omega$ constant for all points on the grid). Otherwise, for a non-uniformly sampled grid the inner product $l_2$ can be replaced by a weighted one $\mathbf{K} = \mathbf{U}^T \mathbf{W} \mathbf{U}$, where $\mathbf{W}$ is a diagonal matrix containing weighting the volumes or areas in the non-uniformly sampled mesh \cite{Scott}.}

POD algorithms for gridded data use Eq.~\eqref{Key_2}. The interpolation-based formulations seek to bring the scattered data onto a uniform grid so that Eq.~\eqref{Key_2} can still be used. We here propose to use the original version in Eq.~\eqref{Key}. Using the regression in Eq.~\eqref{func} from step 1 \textcolor{black}{it becomes:}

\begin{equation}
\label{Key_approx}
\textcolor{black}{K_{ij} = \frac{1}{|\Omega|} \int_\Omega \bm{\tilde{u}}^T(\bm{x}', \mathbf{t}_i) \bm{\tilde{u}}(\bm{x}', \mathbf{t}_j) \, d\bm{x}'\,.}
\end{equation}

To solve the integral, a quadrature method that interrogates the integrand on specific quadrature points is used. With no loss of generality, we here use Gauss-Legendre quadrature because of its excellent trade-off between accuracy and computational costs. \textcolor{black}{ The Gauss-Legendre quadrature method guarantees exact integration of a polynomial function of degree $2n-1$  defined over the interval $ x \in [-1,1]$  by representing the integral as a weighted sum of function values evaluated at  $n$ nodes. These nodes are chosen to be the roots  $x_{L_i}$  of the Legendre polynomial  $P_n(x)$  of degree  $n$ , which is normalized so that $ P_n(1) = 1$. The $i^{th}$  weight $w_{L_i}$ for each node is given by the formula:}

\begin{equation}
\textcolor{black}{w_{L_i} = \frac{2}{\left(1 - x_{L_i}^2\right) \left(P'_n(x_{L_i})\right)^2}.}
\end{equation}

\textcolor{black}{The selection of Legendre roots as quadrature points enables the pre-determination of both the nodes $x_{L_i}$ and their associated weights  $w_{L_i}$, which are readily available in widely tabulated form for various values of $n$. This approach ensures exact integration up to degree  $2n-1$ by leveraging the orthogonality of the Legendre polynomials. While the quadrature is optimized for functions that can be approximated by a polynomial of degree $2n-1$  within the interval, it can also be applied over other intervals by a straightforward scaling transformation, extending its versatility for functions defined outside  $[-1, 1] $. For readers interested in a detailed mathematical background on this and other integration quadratures applicable to this methodology, we refer to the comprehensive treatment in Ref.~\cite{ralston2001first}.
} It is worth emphasising that quadrature methods based on analytic approximations of the integrand yield higher accuracy compared to the mid-point rule implied in Eq.~\eqref{Key_2}.

\vspace{0.1cm}
\textit{\textbf{Step 3: Computing temporal structures}}

The temporal structure of POD modes are eigenvectors of the temporal correlation matrix. Therefore, given the temporal correlation matrix these reads

\begin{equation}
\mathbf{K} =  \boldsymbol{\Psi} \boldsymbol{\Sigma}^2 \boldsymbol{\Psi}^T,
\label{corr}
\end{equation}
where the matrix $\boldsymbol{\Psi}\in\mathbb{R}^{N_t\times r_c}$ collects the eigenvectors along its columns and $\boldsymbol{\Sigma}$ is a diagonal matrix whose elements $\sigma_i$ represent the mode amplitudes (cf. Eq.~\eqref{POD}). This step is identical to the classic snapshot POD for gridded data. 

\vspace{0.1cm}
\textit{\textbf{Step 4: Computing spatial structures}}

The $r^{th}$ spatial structure $\bm{\phi}_r$ is the result of projecting the dataset $\bm{\tilde{u}}(\bm{x},{t})$ onto the temporal structure $\boldsymbol{\psi}_r$. This project requires an inner product in time. The same discussion in Step 2 now applies to the inner product in the time domain, which reads 

\begin{equation}
\bm{\phi}_r(\bm{x}) = \frac{1}{\sigma_r}\langle \bm{\tilde{u}}(\bm{x}, {t}), \psi_r({t}) \rangle = \frac{1}{\sigma_r T}\int_0^T \bm{\tilde{u}}(\bm{x},{t}') ~\psi_r(t')~d{t'},
\label{phi}
\end{equation}
assuming that the time domain is $t\in [0,T]$. Here both the temporal structures and the regression of the snapshots are available on a discrete set of times $\mathbf{t}\in\mathbb{R}^{N_t}$ and should be converted to functions via interpolation or regression. The integral can be computed using the same Gauss-Legendre quadratures as in the definition of $\bm{K}$ in Step 2. However, the gain in accuracy was deemed not sufficient to justify the increase in computational cost. Therefore, in this work, this projection is carried out via mid-point approximation in time\textcolor{black}{:}
\begin{equation}
\textcolor{black}{\frac{1}{\sigma_r T}\int_0^T \bm{\tilde{u}}(\bm{x},{t}) ~\psi_r({t})~d{t} \approx \frac{1}{\sigma_r N_t}\sum^{N_t}_{k=1}\tilde{u}(\mathbf{X},\mathbf{t}_k)\psi_r(\mathbf{t}_k).}
\label{eq.phi_app}
\end{equation}

We stress that Eq.~\eqref{phi} holds for \emph{any} set of points $\bm{x}$. Hence, Eq.~\eqref{phi} enables super-resolution of the POD modes. In this work, we compute it on a uniform grid for plotting purposes and compare it with interpolation-based approaches.

\vspace{0.5cm}

\textcolor{black}{The main novelty in the proposed approach is to replace the discrete $l_2$ inner product with the continuous counterpart leveraging the RBF regression. A number of linear decompositions in modal analysis could be rendered meshless in the same manner. This is the case of the Spectral PODs, the Multiscale POD or the Dynamic Mode Decomposition (see \cite{mendez_2023_chap8} for a review and comparison).}

\textcolor{black}{The proposed approach is termed mesh-less because it does not rely on a grid, either for data representation or for computing the required inner products. Instead, both the regression of the snapshots in Step 1 and the integral computation in Step 2 are performed using collocation methods. The distinction between mesh-based and collocation-based (mesh-free) numerical techniques is discussed in Refs. \cite{Liu2005, Zhang2024}. The regression step offers flexibility in adapting to complex geometries, while the integration process can be replaced by more easily parallelizable methods, such as Monte Carlo methods.}

\begin{algorithm}
\caption{Meshless POD \vspace{0.1cm}}
\begin{algorithmic}[1]
	\Require Data: $\{\bm{u}(\mathbf{X}^{(i)}, \mathbf{t}_i)\}$ for $i = 1, 2, \ldots, N_t$ (scattered in space and time).
	\Ensure POD meshless modes: $\{\sigma, \bm{\phi}(\bm{x}), \psi({t})\}$.
	\vspace{0.1cm}
	\Statex\hrulefill
	\Statex \textit{\textbf{Step 1:} Analytical snapshot representation with RBF}
	\Require Data: $\{\bm{u}(\mathbf{X}^{(i)}, \mathbf{t}_i)\}$ for $i = 1, 2, \ldots, N_t$; number of Gauss-Legendre quadrature points $n$; time vector $\mathbf{t} \in \mathbb{R}^{N_t}$.
	\Ensure $\bm{\tilde{u}}(\mathbf{X}_{L},\mathbf{t}_i)$. 
	\vspace{0.1cm}
	\State Compute and scale the $n$ quadrature points $\mathbf{X}_{L}$ 
	\For{each time step $\mathbf{t}_i$}
	\State Compute analytical approximation through RBF as in Eq.~\eqref{func} 
	\State Interpolate $\{\bm{u}(\mathbf{X}^{(i)}, \mathbf{t}_i)\}$ on $\mathbf{X}_{L}$ and store $\rightarrow$ $\bm{\tilde{u}}(\mathbf{X}_{L},\mathbf{t}_i)$ 
	\EndFor
	\Statex\hrulefill
	\Statex \textit{\textbf{Step 2:} Temporal correlation matrix}
	\Require $\bm{\tilde{u}}(\mathbf{X}_{L},\mathbf{t}_i)$.
	\Ensure Temporal correlation matrix: $\mathbf{K} \in \mathbb{R}^{N_t \times N_t}$. 
	\vspace{0.1cm}
	\State Initialize  $\mathbf{K}$
	\For{each pair of time steps $(\mathbf{t}_i, \mathbf{t}_j)$}
	\State Compute inner product in Eq.~\eqref{K} using Gauss-Legendre quadrature: 
	\[
	K_{ij} = \frac{1}{2^d} \sum_{k=1}^{n} \bm{w}_{L_k}{\bm{\tilde{u}}}(\bm{X}_{L_k},{t_i}) {\bm{\tilde{u}}}(\mathbf{X}_{L_k},{t_j}) 
	\]
	with $\bm{w}_{L}$ the tabulated weights associated with the values of any function on the $\mathbf{X}_{L}$ points and $d$ the dimensionality of the problem.
	\State Store the element $K_{ij}$
	\EndFor
	\Statex\hrulefill
	\Statex \textit{\textbf{Step 3:} Computing temporal structures}
	\Require $\mathbf{K}$. 
	\Ensure $\boldsymbol{\Psi}$ and $\boldsymbol{\Sigma}$. 
	\State Decompose the temporal correlation $\textbf{K}$ trough SVD as in Eq.~\eqref{corr} 
	\State Store $\boldsymbol{\Psi}$ and $\boldsymbol{\Sigma}$
	\Statex\hrulefill
	\Statex \textit{\textbf{Step 4:} Compute spatial structures}
	\Require Data: $\{\bm{u}(\mathbf{X}^{(i)}, \mathbf{t}_i)\}$; output grid $\mathbf{X}_{out}$ (only for visualization purposes) and $\boldsymbol{\Psi}$.  
	\Ensure $\boldsymbol{\Phi}$. 
	\For{each time instant}
	\State Project $\bm{\tilde{u}}(\mathbf{X}^{(i)}, \mathbf{t}_i)$ on $\mathbf{X}_{out} \rightarrow \bm{\tilde{u}}(\mathbf{X}_{out}, \mathbf{t}_i)$: 
	\EndFor   
	\For{r = 1 to rank} 
	\State Compute the inner product in Eq.\eqref{phi}
	with the approximation: 
	\[
	\bm{\phi}_r(\bm{x}) \approx \frac{1}{\sigma_r N_t}\sum^{N_t}_{k=1}\tilde{u}(\mathbf{X_{out}},\mathbf{t}_k)\psi_r(\mathbf{t}_k)
	\]
	\EndFor
	\State Store $\phi_r$ for each $r$
	
\end{algorithmic}
\label{pseudo}
\end{algorithm}

\subsection{Relation to parametric and functional analysis}\label{subsec:parametric_functional}

\textcolor{black}{Extensions of the proposed framework to parametric analysis is straightforward. The proposed formulation focuses on the case of data distributed in space and time, but it can be easily generalized to a parametric setting, where the data is assumed to depend on a variable (say $x$) and a set of parameters $\bm{\mu}\in\mathbb{N_\mu}$. In this case, all the steps from $1$ to $3$ remains unaltered, while the integrals in step $4$ should be replaced by summations as in the traditional POD/PCA.}

\textcolor{black}{Our meshless POD also shares some features with FPCA. This extends PCA to functional data, where each observation consists of a smooth function over a continuous domain (typically time) rather than a discrete set of measurements. These functions are typically constructed using B-splines, Fourier series, or other basis functions, meaning Step 1 in the proposed algorithm is inherently completed. The key differences, however, lie in the subsequent steps. First, functional PCA extends traditional PCA (or classic POD) rather than dual PCA (or snapshot POD) as used in the current method. Second, since the "snapshots" in FPCA consist of continuous functions, the correlation matrix is replaced by a correlation function, turning the eigenvalue problem into a continuous one. For the types of problems considered in this work, this results in a significantly more computationally intensive task. To be more specific, implementing FPCA in our context would proceed as follows. Let $\bm{u}(\bm{x},t_k)=\bm{\Gamma}(\bm{x})\bm{w}_k$ represent the RBF regression of a snapshot at a discrete time $t_k$, where $\bm{\Gamma}(\bm{x}) \in \mathbb{R}^{n_x \times n_b}$ is the matrix of the $n_b$ RBFs evaluated at an arbitrary set of $n_x$ points in $\Omega$, and $\bm{w}_k \in \mathbb{R}^{n_b}$ is the corresponding set of weights. The correlation function and the resulting eigenvalue problem can then be expressed as}

\begin{equation}
C(\bm{x},\bm{x}')=\frac{1}{N_t} \sum^{N_t}_{k=1} \bm{u}(\bm{x},t_k)\bm{u}(\bm{x}',t_k) \quad \mbox{and}\quad \int_{\Omega}C(\bm{x},\bm{x}') \bm{\phi}_n d\Omega = \bm{\phi}_n(\bm{x}) \lambda_n\,.
\end{equation}

\textcolor{black}{If the RBF method is used to approximate the integral eigenvalue problem, it reduces to diagonalizing a matrix of size $n_b \times n_b$. An RBF approximation of the eigenfunction $\phi_n(\bm{x})$ is then obtained on the same basis $\Gamma(\bm{x})$ used for interpolating the snapshots. In contrast, the eigenvalue problem in the proposed approach has a size of $N_t \times N_t$ and is formulated as a discrete problem from the outset (step 3).}

\section{Validation}\label{Sec.validation}
\subsection{Analytical testcase}\label{sub.sine}
\begin{figure}
\centering
\includegraphics{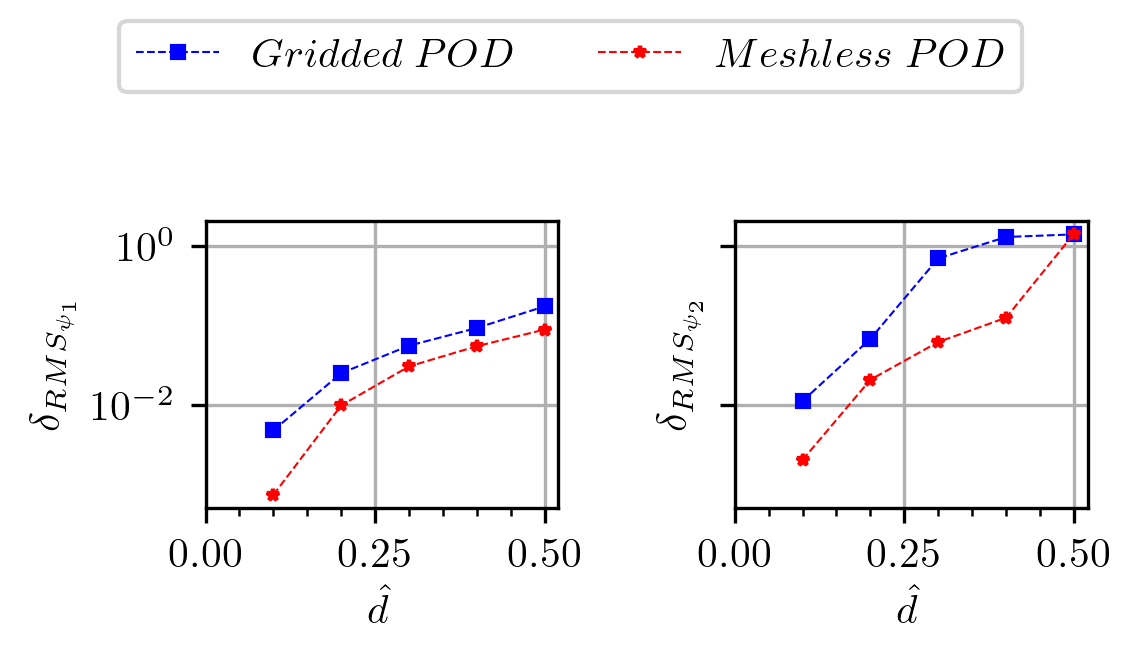}
\caption{Root mean square error of the $i^{th}$ temporal mode $\psi_i$ ($\delta_{RMS_{\psi_i}}$) as function of  $\hat{d}$ for the analytical testcase. The values are normalised with the standard deviation of its corresponding reference modes $\psi_{ref_i}$. The red curve with star markers depicts the meshless POD, while the blue
	curve with square markers represents the gridded POD.}
\label{fig:temp_analytical}
\end{figure}

\begin{figure}[t]
\centering
\begin{overpic}[scale=1,unit=1mm]{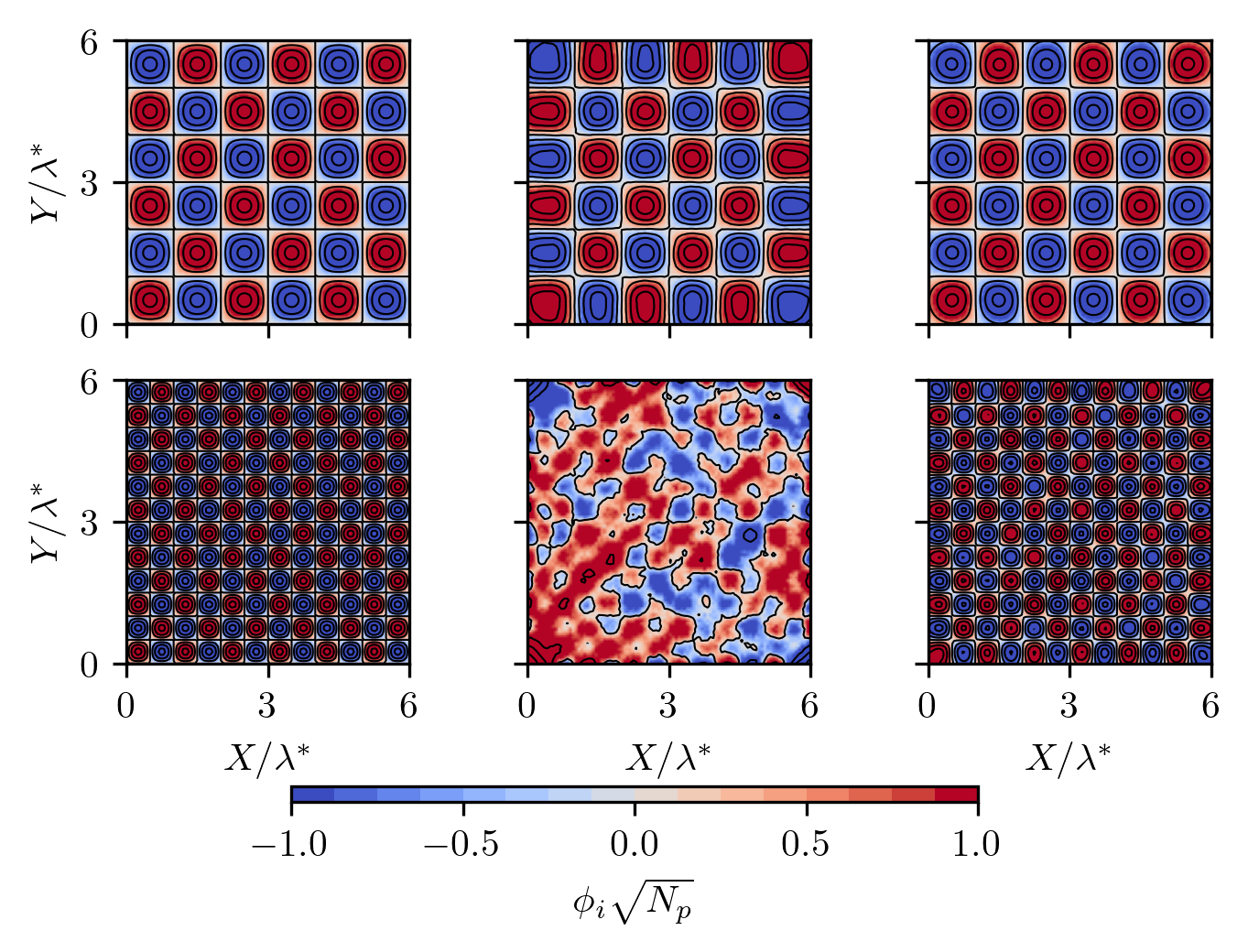}
	
	\put(11,88){\parbox{25mm}{\centering \textnormal{{Reference POD}}}}
	\put(49,88){\parbox{25mm}{\centering \textnormal{{Gridded POD}}}}
	\put(86,88){\parbox{25mm}{\centering \textnormal{{Meshless POD}}}}
	\put(0,0){\parbox{10mm}{\centering \textnormal{{b)}}}}
	\put(0,92){\parbox{10mm}{\centering \textnormal{{a)}}}}
	
\end{overpic}
\centering

\includegraphics[scale = 1]{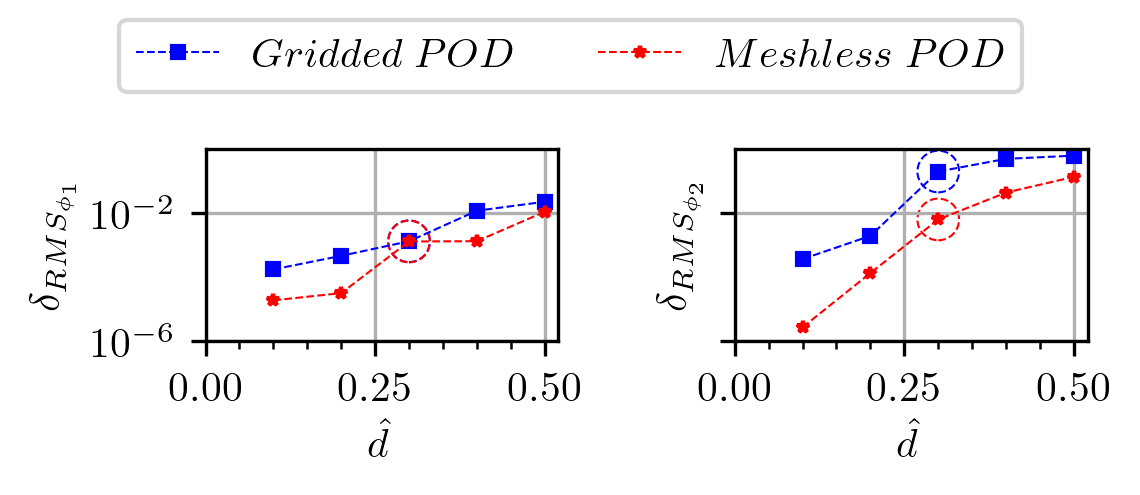}
\caption{a) Comparison of the spatial modes $\phi_i$  for the analytical test case. The reference POD is represented in the first column, while the gridded and the meshless approaches are displayed in the second and the third columns, respectively. Only the first two modes are pictured. Case $\hat{d} =$ \textcolor{black}{$0.3$}  b) Root mean square error of the first two spatial modes $\phi$  ($\delta_{RMS_{\phi_i}}$) as a function of  $\hat{d}$.  The values are normalised with the standard deviation of its corresponding reference modes $\phi_{ref_i}$. The dashed circles highlight the case pictured on the top part of the figure. }
\label{fig:spat_analyt}
\end{figure}
A synthetic test case was first considered for benchmarking the proposed method on a problem for which the POD is analytically available. This allows for isolating all sources of errors. In this test case, we consider a spatial domain of length $L_x$ and $L_y$ that spans $x,y \in [0,12]$, discretised with $100$ grid points. Two 2D sinusoidal orthogonal waves have been chosen as spatial structures, with a leading wavenumber $k_x=k_y = 3$ in both $x$ and $y$ direction:

\begin{equation}
\label{Phi_P}
\phi_1 (x,y) = \sin\left( k_x x\right) \sin\left( k_y y\right) \,,\,\phi_2 (x,y) = \sin\left(2k_x x\right) \sin\left(2 k_y y\right).
\end{equation}

Similarly, the associated temporal structures are sampled over a period $T = 10$ discretised with $500$ time instants: 

\begin{equation}
\label{Psi_P}
\psi_1 = \sin\left(\frac{4\pi}{T} \cdot t\right) \,,\,
\psi_2 = \sin\left(\frac{2\pi}{T} t\right). 
\end{equation}

Assigning an amplitude $\sigma_1=10$ and $\sigma_2=5$ to each mode respectively, the synthetic dataset $\mathbf{D}$ is built as a linear combination of these structures:

\begin{equation}
\mathbf{D} = \sigma_1\phi_1\psi_1^T + \sigma_2\phi_2\psi_2^T ,  
\end{equation}

\noindent whose analytical modes represent the ground truth. It is easy to show that the modes with spatial structures \eqref{Phi_P} and temporal structures \eqref{Psi_P} are indeed POD modes, being orthogonal both in space and time according to the standard inner product. We consider them as \textbf{``reference POD"}.

A parametric study involving the influence of the sparsity of data has been carried out. To this end, we define the parameter $\hat{d}$ to refer to the different cases, which represents the average distance among the ``sensors" normalised with the reference lengthscale $l_{ref}$:
\begin{equation}
\hat{d} = \frac{l_{ref}}{\sqrt{N_{ppp}}},
\label{eq.d}
\end{equation}

\noindent where $N_{ppp}$ represents the sensor density. In this specific application, the reference length-scale is the smallest achievable wavelength $\lambda^* = \frac{L_x}{2k_x}$.

The different datasets are obtained for $0.1 <\hat{d} < 0.5$, whose corresponding number of particles is computed inverting Eq.~\eqref{eq.d}. These datasets serve as test cases for the meshless POD computation, and the corresponding results are denoted as \textcolor{black}{``meshless POD''}. The mapping onto a regular grid is performed using a moving average, with bin sizes determined ensuring approximately $10$ sensor within each bin. This is a common practice in particle-image-based measurements, for instance, and introduces a high degree of robustness to noise and outliers, although at the expenses of spatial resolution. From this mapping, the classic matrix-factorisation approach for the POD is employed. In what follows, we refer to the results of this approach on the gridded data as \textcolor{black}{``gridded POD"}.
\textcolor{black}{While the binning approach does not preserve energy, the same applies to the RBF-based method, as it provides an approximation where the integral is not solved exactly. However, it is important to note that this is not an inherent limitation of either algorithm. Since these methods are purely data-driven, there is no guarantee that the data itself captures all the energy in the flow. In any case, no constraint on energy preservation is imposed, ensuring a fair comparison among the methodologies.} 

Figure~\ref{fig:temp_analytical} illustrates the root mean square error of the temporal structures $\psi_i$ ($\delta_{RMS_{\psi_i}}$), normalised by the standard deviation of the corresponding
reference modes, for different values of $\hat{d}$,  \textcolor{black}{computed as:}
\begin{equation}
\textcolor{black}{\delta_{RMS_{\psi_i}} = \frac{\sqrt{\frac{1}{N_t} \sum_{j=1}^{N_t} \left( \psi_{i,j} - \psi_{\text{ref}_ {i,j}} \right)^2}}{\sqrt{\frac{1}{N_t} \sum_{j=1}^{N_t} \psi_{\text{ref}_{i,j}}^2}}}.
\end{equation}

This test highlights the superior accuracy of the meshless approach compared to the traditional one, even for sparser datasets. A similar trend is observed for the spatial structures depicted in Fig.\ref{fig:spat_analyt}.

Fig.~\ref{fig:spat_analyt}.a presents a visual comparison for the case $\hat{d} = 0.3$ between the reconstructed spatial structures and the analytical ones. \textcolor{black}{The meshless modes are obtained trough Eq.~\eqref{eq.phi_app}, while the reference and the gridded trough classic eigenvalue decomposition.} It is evident from this figure, particularly in the second mode, that the gridded approach filters out smaller scales more aggressively. This observation is further confirmed by the plot shown in Fig.~\ref{fig:spat_analyt}.b. As for the temporal case, the $\delta_{RMS_{\phi_i}}$ is normalised with the standard deviation of the corresponding reference modes:

\begin{equation}
\textcolor{black}{\delta_{RMS_{\phi_i}} = \frac{\sqrt{\frac{1}{N_p} \sum_{j=1}^{N_p} \left( \phi_{i,j} - \phi_{\text{ref}_ {i,j}} \right)^2}}{\sqrt{\frac{1}{N_p} \sum_{j=1}^{N_p} \phi_{\text{ref}_{i,j}}^2}}}.  
\end{equation}

The benchmark on this toy problem, despite its simplicity, enabled us to eliminate external sources of error in the data originating from simulations or experiments. This process allowed us to evaluate the suitability of the proposed methodology when compared with the traditional approach.

\subsection{Fluidic pinball}\label{Sec.pinball}
\begin{figure}[t]
\centering
\includegraphics{{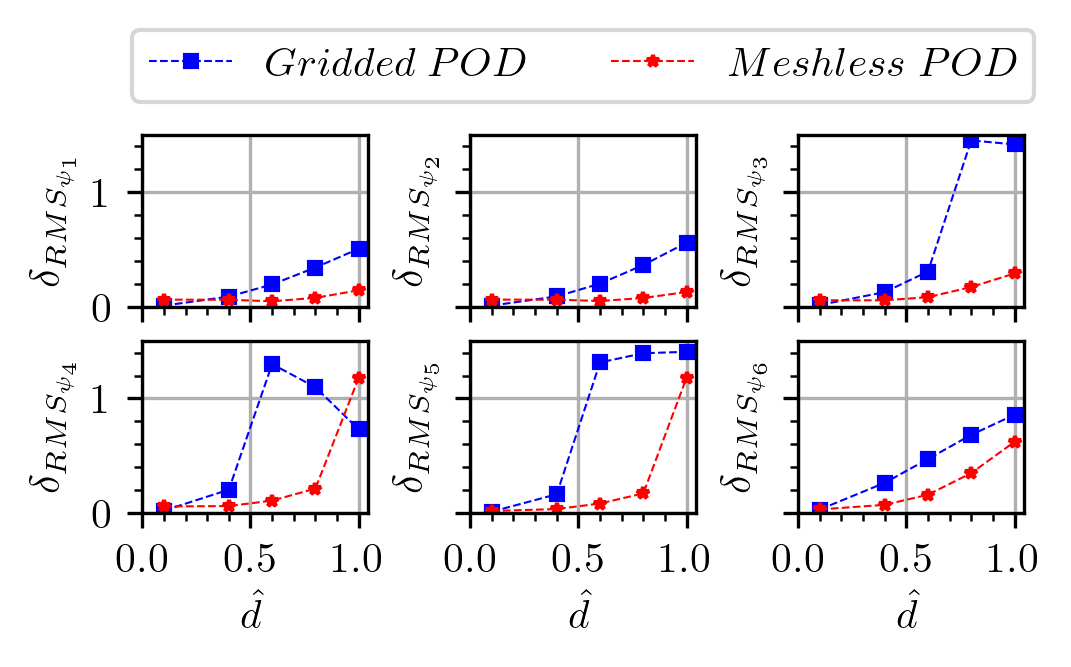}}

\caption{ Root mean square error of the $i^{th}$ temporal mode $\psi_i$  ($\delta_{RMS_{\psi_i}}$) as function of  $\hat{d}$ for the fludic pinball testcase. The values are normalised with the standard deviation of its corresponding reference modes $\psi_{ref_i}$. The red curve with star markers depicts the meshless POD, while the blue
	curve with square markers represents the gridded POD.}
\label{fig:temp_pin}
\end{figure}

\begin{figure}[t]
\centering
\begin{overpic}[scale=1,unit=1mm]{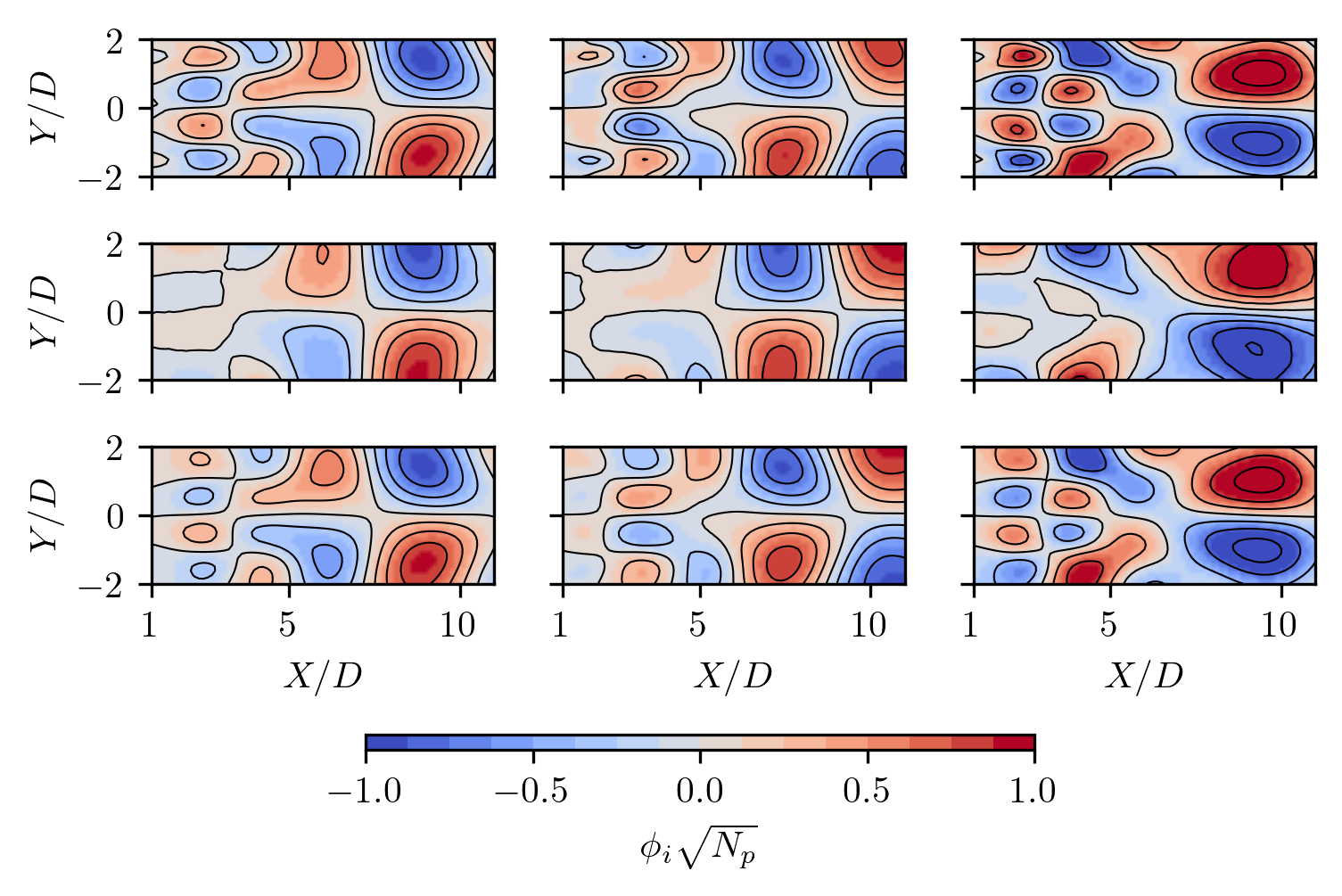}
	
	\put(50,83){\parbox{40mm}{\centering \textnormal{{Reference POD}}}}
	\put(50,64){\parbox{40mm}{\centering \textnormal{{Gridded POD}}}}
	\put(50,45){\parbox{40mm}{\centering \textnormal{{Meshless POD}}}}
	\put(0,0){\parbox{10mm}{\centering \textnormal{{b)}}}}
	\put(0,85){\parbox{10mm}{\centering \textnormal{{a)}}}}
	
\end{overpic}
\centering

\includegraphics[scale = 1]{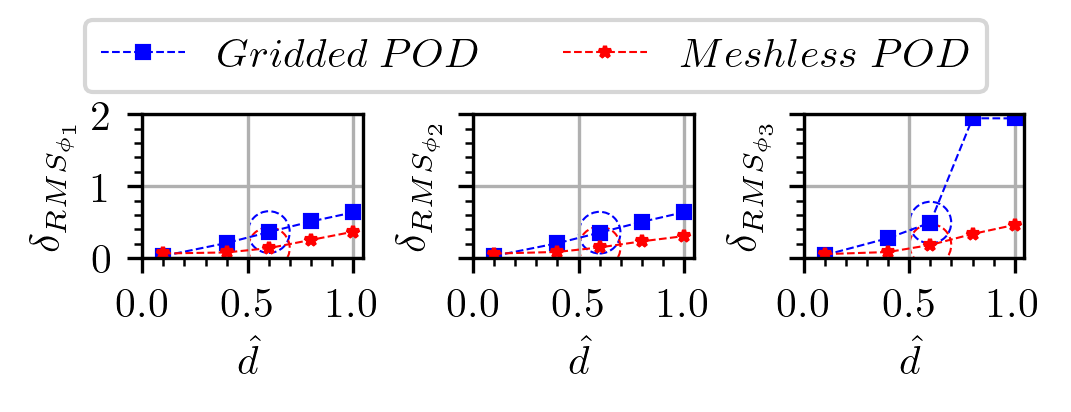}
\caption{a) Comparison of the spatial modes $\phi_i$ associated to the streamwise velocity component $u$ for the fluidic pinball. The reference POD is represented in the first row, while the gridded and the meshless approaches are displayed in the second and the third rows respectively. Only the first three modes are pictured. Case $\hat{d} = 0.6$  b) Root mean square error of the first three spatial modes $\phi_i$  ($\delta_{RMS_{\phi_i}}$) as a function of  $\hat{d}$.  The values are normalised with the standard deviation of its corresponding reference modes $\phi_{ref_i}$. The dashed circles highlight the case pictured on the top part of the figure. The red curve with star markers depicts the meshless approach, while the blue curve with square markers represents the gridded method. }
\label{fig:spat_pin}
\end{figure}

The wake of a fluidic pinball is a popular test case in fluid mechanics. A pinball consists of a configuration of three cylinders with diameter $D$, with axes positioned at the vertices of an equilateral triangle with side $3D/2$. The pinball is invested by a uniform constant wind of velocity $U_\infty$.
Direct Numerical Simulation (DNS) data from Ref.~\cite{deng2020low} is used for this case. The DNS data consists of an unstructured grid with $3536$ points within the domain $x/D \in [1, 11]$ and $y/D \in [-2, 2]$ (with $x,y$ being the stream-wise and cross-wise directions, respectively). The triangle formed by the cylinder axis points upstream, with the downstream edge orthogonal to the mean flow direction. The midpoint of the downstream edge is located at $(0,0)$. The flow kinematic viscosity $\nu$ is set to have a Reynolds number $Re=U_\infty D/\nu=150$, in which the flow exhibits a chaotic behaviour \cite{deng2020low}.

We reproduce the standard configuration of Particle Tracking Velocimetry (PTV) measurements. To this purpose, it is assumed that the domain is observed by a camera with a resolution of $32$ pixels per diameter. The velocity fields are randomly sampled (as if they were originated from particle measurements) at different concentrations, resulting in a mean distance in the range  $0.1 < \hat{d} < 1$ \textcolor{black}{( or in terms of total number of particles $3536 > N_p > 40$)}. The cylinder diameter $D$ has been used here as reference lengthscale. A baseline for the POD computation is obtained by \textcolor{black}{a direct interpolation with RBF of the DNS data on a fine grid with spacing $\Delta x=D/16$,} \textcolor{black}{ while the computation of the POD with the traditional matrix factorisation and the meshless approach is carried out on these downsampled distributions}.

Since the dataset is statistically stationary, focus is placed on the decomposition of the fluctuating component of the velocity field. To ensure a fair comparison across the methodologies, an ensemble high-resolution mean is subtracted from all the datasets as explained in Ref.~\cite{tirelli2023simple}. 

Fig.~\ref{fig:temp_pin} shows the root mean square error of the first six temporal structures $\psi_i$ ($\delta_{RMS_{\psi_i}}$), normalised with the square root of the number of samples $N_t$, for different $\hat{d}$. 
This kind of comparison is limited to the most energetic modes. Higher-order modes have generally smaller energy difference, which might result in different sorting for different decomposition methods. The illustrated modes suggest that the meshless approach closely follows the reference. 
The overall trend indicates that an increase in the average distance among sensors corresponds to an increase in the error. Notably, the gridded approach is more susceptible to this effect. To accommodate the larger distances, the method tends to augment the bin size to ensure a sufficient number of sampling points within the bin. Nonetheless, this strategy unintentionally filters out smaller scales and exacerbates spatial resolution limitations. On the other hand, the meshless regression appears more robust to the detrimental effect of increasing the sparsity of the sensors.  It is worth remarking that values of $\delta_{RMS_{\psi_i}} > 1$ might also be originating from the comparison among temporal structures that are not synchronised with the reference ones due to different sorting of modes with similar energy, as discussed before.

The first $3$ spatial modes $\phi_i$ associated with the stream-wise velocity component $U$ are analysed in Fig.~\ref{fig:spat_pin}.a. The contours in the top part of this figure are representative of the case $\hat{d} = 0.6$, as highlighted by the dashed circles in the bottom part of the picture. These contours show that the gridded POD exhibits more attenuated modes if compared to the ones computed with the meshless approach. In Fig.~\ref{fig:spat_pin}.b the $\delta_{RMS_{\phi_i}}$ of these modes normalised with the rms of the corresponding reference modes is reported as a function of $\hat{d}$. Similarly to the temporal modes, the meshless approach shows superior robustness to increased data sparsity.

\subsection{ C3S precipitation  from satellite microwave observations }\label{sub.meteo}
\begin{figure}[t]
\centering
\includegraphics{{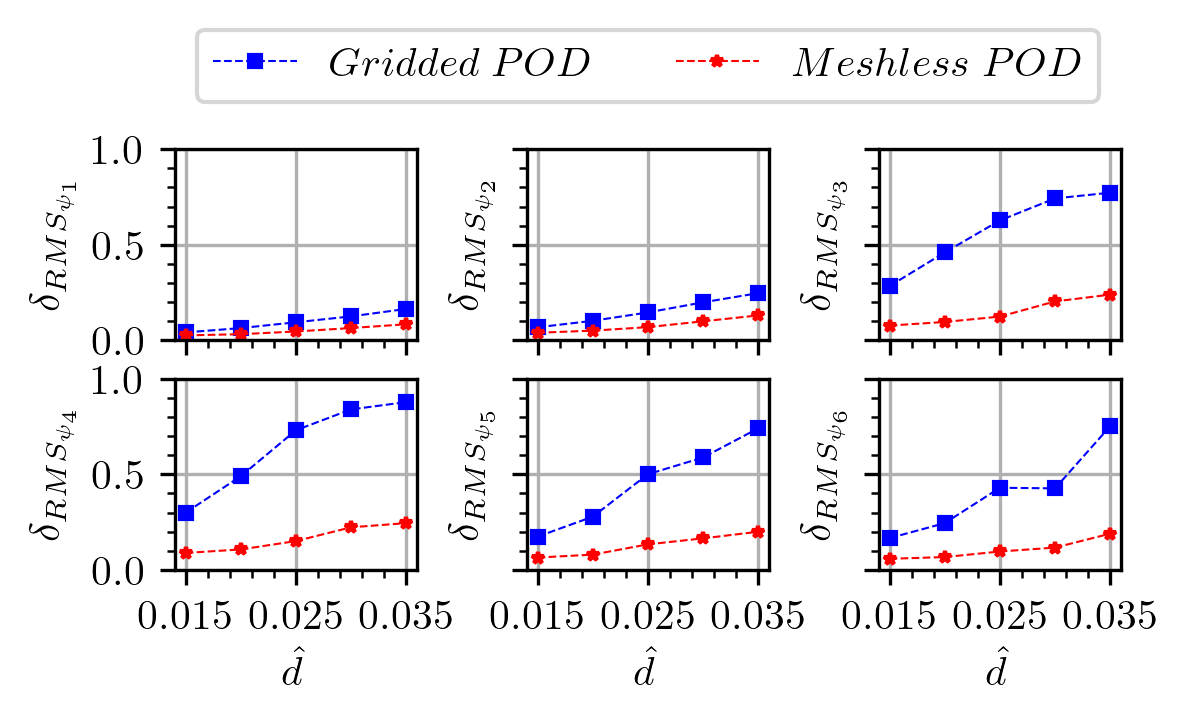}}

\caption{ Root mean square error of the $i^{th}$ temporal mode $\psi_i$  ($\delta_{RMS_{\psi_i}}$) as function of  $\hat{d}$ for the C3S precipitation testcase. The values are normalised with the standard deviation of its corresponding reference modes $\psi_{ref_i}$. The red curve with star markers depicts the meshless POD, while the blue
	curve with square markers represents the gridded POD.}
\label{fig:temp_prep}
\end{figure}

\begin{figure}[t]
\centering
\begin{overpic}[scale=1,unit=1mm]{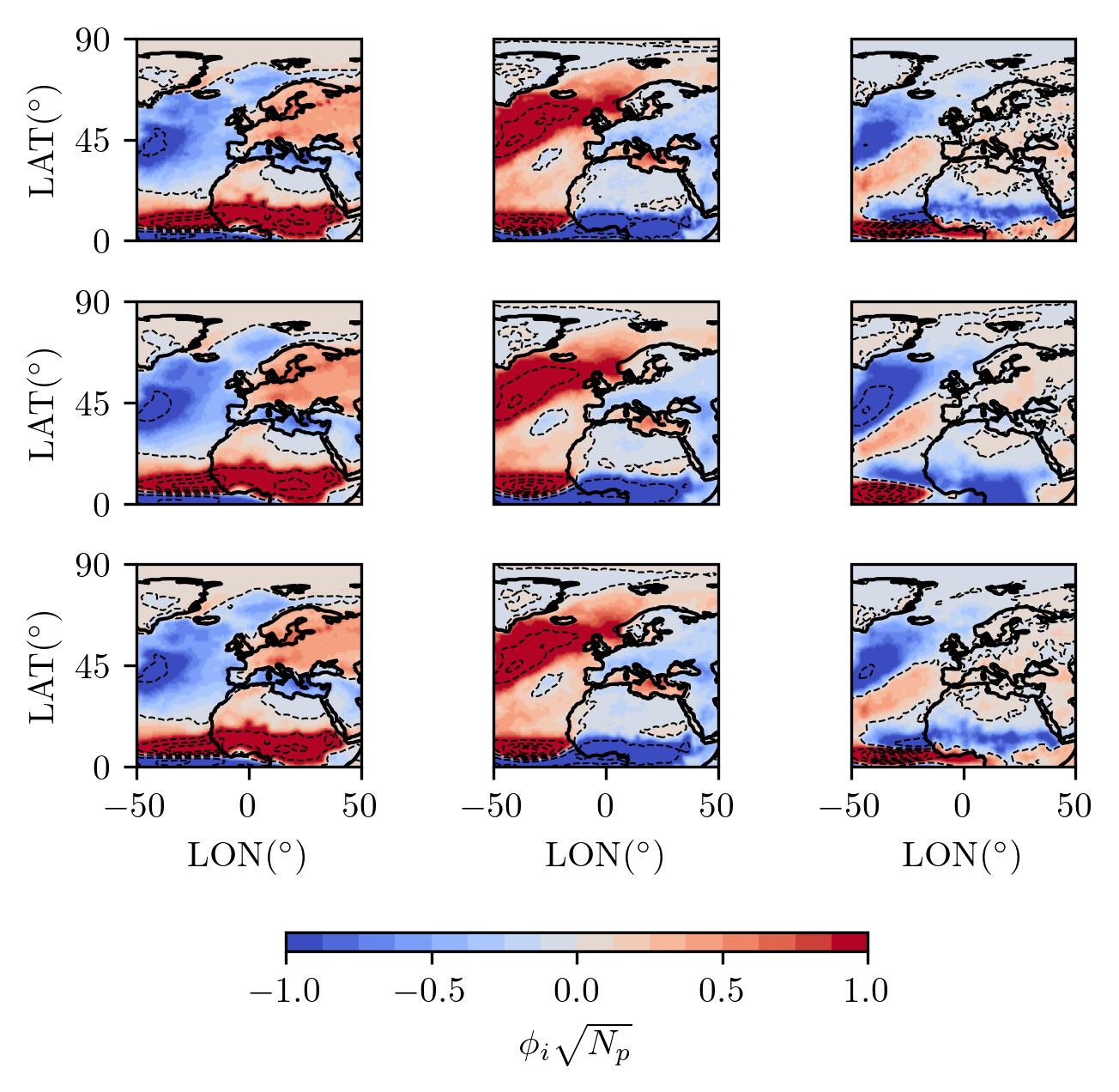}
	
	\put(40,105){\parbox{40mm}{\centering \textnormal{{Reference POD}}}}
	\put(40,79){\parbox{40mm}{\centering \textnormal{{Gridded POD}}}}
	\put(40,53.5){\parbox{40mm}{\centering \textnormal{{Meshless POD}}}}
	\put(0,0){\parbox{10mm}{\centering \textnormal{{b)}}}}
	\put(0,106){\parbox{10mm}{\centering \textnormal{{a)}}}}
	
\end{overpic}
\centering

\includegraphics[scale = 1]{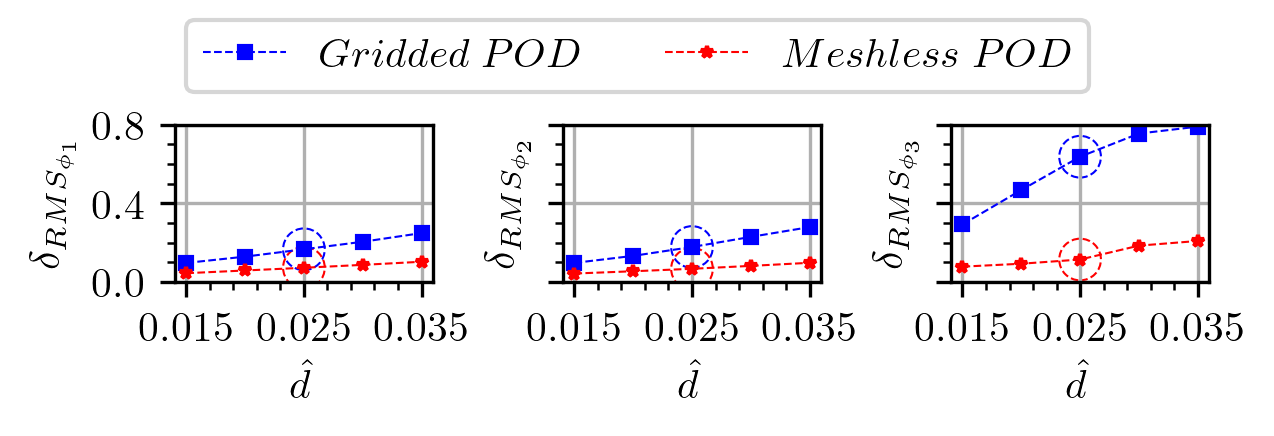}
\caption{a) Comparison of the spatial modes $\phi_i$ associated to the anomaly of daily precipitation. The reference POD is represented in the first row, while the gridded and the meshless approach are displayed in the second and the third rows respectively. Only the first three modes are pictured. Case $\hat{d} = 0.025$  b) Root mean square error of the first three spatial modes $\phi_i$  ($\delta_{RMS_{\phi_i}}$) as a function of  $\hat{d}$.  The values are normalised with the standard deviation of its corresponding reference modes $\phi_{ref_i}$.  }
\label{fig:spat_prep}
\end{figure}
The experimental testcase is based on the  ``Copernicus Climate Change Service (C3S) Climate Data Store (CDS): Precipitation monthly and daily gridded data from 2000 to 2017 derived from satellite microwave observations''\citep{Konread}.
\textcolor{black}{The choice of this third dataset aims to broaden the scope of this work beyond the fluid dynamics field, demonstrating that this methodology is well-suited for more general applications with scattered data, as the database of meteorological data presented here.}

This dataset provides global estimates of daily accumulated precipitation and monthly mean precipitation. These estimates are derived from a combination of passive microwave observations obtained from two different types of radiometers on various Low Earth Orbit satellites. Firstly, there are the conically scanning microwave imagers, which undergo processing utilising methodologies established by the Hamburg Ocean Atmosphere Parameters and Fluxes from Satellite initiative, as part of the Satellite Application Facility on Climate Monitoring.
Secondly, the cross-track scanning microwave sounders whose data undergoes processing through the Passive microwave Neural network Precipitation Retrieval for Climate Applications algorithm.

The main variable in this dataset is the precipitation $\mathcal{P}$, expressed in $mm/day$, which represents the water-equivalent volume rate per area and per day of atmospheric water in liquid or solid phase reaching the Earth's surface. The original data are on a uniform grid with a spacing of $1^{\circ}$, where latitudinal and longitudinal coordinates span $[-90^{\circ}, 90^{\circ}]$ and $[-180^{\circ}, 180^{\circ}]$ respectively. The ground-truth for this application is a downsampled version of the original data on a coarser grid with a spacing of $2^{\circ}$, restricted to the region $[-50 ^{\circ},50^{\circ}]$ in longitude (LON) and $[0^{\circ},90^{\circ}]$ in latitude (LAT). The observations are made daily and cover the period from 2000 to 2017, leading to a total of $6576$ days (treated here as snapshots in the POD formulation).

To simulate sparse sensor acquisition from this data, values for $\hat{d} = [0.015,0.02,0.025,0.03,0.035]$ are interpolated for random positions. To compute these values of $\hat{d}$, the latitude and longitude coordinates have been converted into radians, aligning with the reference length scale of the Earth's radius. These sensor distributions are then employed to perform gridded and meshless POD, as in the previous sections. Fig.~\ref{fig:temp_prep} displays the root mean square error of the first $6$ temporal structures, normalised as usual with the square root of the number of snapshots. In Fig.~\ref{fig:spat_prep}, a comparison among the spatial structures for different sparsity levels is proposed. Specifically, Fig.~\ref{fig:spat_prep}.b shows the overall comparison, while Fig.~\ref{fig:spat_prep}.a highlights the specific case of $\hat{d} = 0.02$.

\section{Conclusion}
We propose a novel meshless approach to compute spatial and temporal modes from scattered measurements eliminating the need for interpolation onto a structured fixed grid. Key enablers are RBF regression and advanced quadrature techniques to compute inner products in space and time, providing an analytic (mesh-independent) representation of the modes. By avoiding the modulation effects associated with the unnecessary mapping of scattered measurements onto a Cartesian grid, our method offers advantages in accuracy and efficiency. Furthermore, the quadrature methods enabled by RBF regression allow for more precise computation of the temporal correlation matrix and, consequently, the temporal eigenvectors. We demonstrate that our method enables the recovery of scales filtered out by binning onto a regular grid.

This work serves as the first step towards a broader endeavour. The concepts presented here can be readily extended to the different variants and applications discussed in previous sections. We envisage future developments to incorporate physical constraints in the regression process, extension to a wide range of modal decompositions, and applications of the meshless analytical treatment to method such as Galerkin projections, among others.\vskip6pt

\section*{Acknowledgment}
This project has received funding from the European Research Council (ERC) under the European Union’s Horizon 2020 research and innovation program (grant agreement No 949085). Views and opinions expressed are however those of the authors only and do not necessarily reflect those of the European Union or the European Research Council. Neither the European Union nor the granting authority can be held responsible for them.

\bibliography{sn-bibliography}

\begin{thebibliography}{99}

\bibitem{Holmes_Lumley_Berkooz_1996}
Holmes P, Lumley JL, Berkooz G. 1996 {\em Turbulence, Coherent Structures, Dynamical Systems and Symmetry}.
Cambridge Monographs on Mechanics. Cambridge: Cambridge University Press.
(\href{http://dx.doi.org/10.1017/CBO9780511622700}{10.1017/CBO9780511622700})

\bibitem{Martinson_2018}
Martinson DG. 2018 p. 495–534.
In {\em Empirical Orthogonal Function (EOF) Analysis}, p. 495–534. Cambridge: Cambridge University Press.
(\href{http://dx.doi.org/10.1017/9781139342568.016}{10.1017/9781139342568.016})

\bibitem{Hesthaven2018}
Hesthaven J, Ubbiali S. 2018  Non-intrusive reduced order modeling of nonlinear problems using neural networks. {\em Journal of Computational Physics} \textbf{363}, 55--78.
(\href{http://dx.doi.org/10.1016/j.jcp.2018.02.037}{10.1016/j.jcp.2018.02.037})

\bibitem{Muecke2021}
Mücke NT, Bohté SM, Oosterlee CW. 2021  Reduced order modeling for parameterized time-dependent PDEs using spatially and memory aware deep learning. {\em Journal of Computational Science} \textbf{53}, 101408.
(\href{http://dx.doi.org/10.1016/j.jocs.2021.101408}{10.1016/j.jocs.2021.101408})

\bibitem{Fresca2022}
Fresca S, Manzoni A. 2022  POD-DL-ROM: Enhancing deep learning-based reduced order models for nonlinear parametrized PDEs by proper orthogonal decomposition. {\em Computer Methods in Applied Mechanics and Engineering} \textbf{388}, 114181.
(\href{http://dx.doi.org/10.1016/j.cma.2021.114181}{10.1016/j.cma.2021.114181})

\bibitem{brivio2024ptpi}
Brivio S, Fresca S, Manzoni A. 2024  PTPI-DL-ROMs: pre-trained physics-informed deep learning-based reduced order models for nonlinear parametrized PDEs. {\em arXiv preprint arXiv:2405.08558}.
(\href{http://dx.doi.org/10.48550/arXiv.2405.08558}{10.48550/arXiv.2405.08558})

\bibitem{mendez_2023_chap8}
Mendez M. 2023  Generalized and Multiscale Modal Analysis. In Mendez MA, Ianiro A, Noack BR, Brunton SL, editors, {\em Data-Driven Fluid Mechanics: Combining First Principles and Machine Learning} ,  p. 153–181. Cambridge: Cambridge University Press.
(\href{http://dx.doi.org/10.1017/9781108896214.013}{10.1017/9781108896214.013})

\bibitem{kirby1990application}
Kirby M, Sirovich L. 1990  Application of the Karhunen-Loeve procedure for the characterization of human faces. {\em IEEE Transactions on Pattern analysis and Machine intelligence} \textbf{12}, 103--108.
(\href{http://dx.doi.org/10.1109/34.41390}{10.1109/34.41390})

\bibitem{saini2016development}
Saini P, Arndt CM, Steinberg AM. 2016  Development and evaluation of gappy-{POD} as a data reconstruction technique for noisy {PIV} measurements in gas turbine combustors. {\em Experiments in Fluids} \textbf{57}, 1--15.
(\href{http://dx.doi.org/10.1007/s00348-016-2208-7}{10.1007/s00348-016-2208-7})

\bibitem{yao2017empirical}
Yao Z, Wang Z, Forrest JYL, Wang Q, Lv J. 2017  Empirical mode decomposition-adaptive least squares method for dynamic calibration of pressure sensors. {\em Measurement Science and Technology} \textbf{28}, 045010.
(\href{http://dx.doi.org/10.1088/1361-6501/aa5c25}{10.1088/1361-6501/aa5c25})

\bibitem{shen2022informative}
Shen Z, Shi Z, Shen G, Zhen D, Gu F, Ball A. 2022  Informative singular value decomposition and its application in fault detection of planetary gearbox. {\em Measurement Science and Technology} \textbf{33}, 085010.
(\href{http://dx.doi.org/10.1088/1361-6501/ac69b0}{10.1088/1361-6501/ac69b0})

\bibitem{castillo2020data}
Castillo A, Messina AR. 2020  Data-driven sensor placement for state reconstruction via POD analysis. {\em IET Generation, Transmission \& Distribution} \textbf{14}, 656--664.
(\href{http://dx.doi.org/10.1049/iet-gtd.2019.0199}{10.1049/iet-gtd.2019.0199})

\bibitem{pearson1901principal}
Pearson K. 1901  Principal components analysis. {\em The London, Edinburgh, and Dublin Philosophical Magazine and Journal of Science} \textbf{6}, 559.
(\href{http://dx.doi.org/10.1080/14786440109462720}{10.1080/14786440109462720})

\bibitem{hotelling1936simplified}
Hotelling H. 1936  Simplified calculation of principal components. {\em Psychometrika} \textbf{1}, 27--35.
(\href{http://dx.doi.org/10.1007/BF02287921}{10.1007/BF02287921})

\bibitem{schmidt1907theorie}
Schmidt E. 1907  Zur Theorie der linearen und nicht linearen Integralgleichungen Zweite Abhandlung: Aufl{\"o}sung der allgemeinen linearen Integralgleichung. {\em Mathematische Annalen} \textbf{64}, 161--174.
(\href{http://dx.doi.org/10.1007/BF01449890}{10.1007/BF01449890})

\bibitem{obukhov1947statistically}
Obukhov AM. 1947  Statistically homogeneous fields on a sphere. {\em Usp. Mat. Nauk} \textbf{2}, 196--198.
(\href{http://dx.doi.org/10.1016/S0924-7963(02)00240-3}{10.1016/S0924-7963(02)00240-3})

\bibitem{lorenz1956empirical}
Lorenz EN. 1956 {\em Empirical orthogonal functions and statistical weather prediction} vol.~1.
Cambridge: Massachusetts Institute of Technology, Department of Meteorology Cambridge.

\bibitem{kutzbach1967empirical}
Kutzbach JE. 1967  Empirical eigenvectors of sea-level pressure, surface temperature and precipitation complexes over North America. {\em Journal of Applied Meteorology and Climatology} \textbf{6}, 791--802.
(\href{http://dx.doi.org/10.1175/1520-0450(1967)006<0791:EEOSLP>2.0.CO;2}{10.1175/1520-0450(1967)006<0791:EEOSLP>2.0.CO;2})

\bibitem{lumley1967structure}
Lumley JL. 1967  The structure of inhomogeneous turbulent flows. {\em Atmospheric turbulence and radio wave propagation} pp. 166--178.

\bibitem{sirovich1987turbulence}
Sirovich L. 1987  Turbulence and the dynamics of coherent structures. I. Coherent structures. {\em Quarterly of applied mathematics} \textbf{45}, 561--571.
(\href{http://dx.doi.org/10.1090/qam/910463}{10.1090/qam/910463})

\bibitem{sirovich1991analysis}
Sirovich L. 1991  Analysis of turbulent flows by means of the empirical eigenfunctions. {\em Fluid Dynamics Research} \textbf{8}, 85.
(\href{http://dx.doi.org/10.1016/0169-5983(91)90033-F}{10.1016/0169-5983(91)90033-F})

\bibitem{Ghojogh2019}
Ghojogh B, Crowley M. 2019  Unsupervised and Supervised Principal Component Analysis: Tutorial. (\href{http://dx.doi.org/10.48550/ARXIV.1906.03148}{10.48550/ARXIV.1906.03148})

\bibitem{Ghojogh2019a}
Ghojogh B, Samad MN, Mashhadi SA, Kapoor T, Ali W, Karray F, Crowley M. 2019  Feature Selection and Feature Extraction in Pattern Analysis: A Literature Review. (\href{http://dx.doi.org/10.48550/ARXIV.1905.02845}{10.48550/ARXIV.1905.02845})

\bibitem{Schoelkopf1997}
Schölkopf B, Smola A, Müller KR. 1997 pp. 583--588.
In {\em Kernel principal component analysis}, pp. 583--588. Berlin, Heidelberg: Springer Berlin Heidelberg.
(\href{http://dx.doi.org/10.1007/bfb0020217}{10.1007/bfb0020217})

\bibitem{Jolliffe2016}
Jolliffe IT, Cadima J. 2016  Principal component analysis: a review and recent developments. {\em Philosophical Transactions of the Royal Society A: Mathematical, Physical and Engineering Sciences} \textbf{374}, 20150202.
(\href{http://dx.doi.org/10.1098/rsta.2015.0202}{10.1098/rsta.2015.0202})

\bibitem{Mendez2023_MST}
Mendez MA. 2023  Linear and nonlinear dimensionality reduction from fluid mechanics to machine learning. {\em Measurement Science and Technology} \textbf{34}, 042001.
(\href{http://dx.doi.org/10.1088/1361-6501/acaffe}{10.1088/1361-6501/acaffe})

\bibitem{jimenez_2023}
Jiménez J. 2023  Coherent Structures in Turbulence. In Mendez MA, Ianiro A, Noack BR, Brunton SL, editors, {\em Data-Driven Fluid Mechanics: Combining First Principles and Machine Learning} ,  p. 20–33. Cambridge: Cambridge University Press.
(\href{http://dx.doi.org/10.1017/9781108896214}{10.1017/9781108896214})

\bibitem{Benner2015}
Benner P, Gugercin S, Willcox K. 2015  A Survey of Projection-Based Model Reduction Methods for Parametric Dynamical Systems. {\em SIAM Review} \textbf{57}, 483--531.
(\href{http://dx.doi.org/10.1137/130932715}{10.1137/130932715})

\bibitem{Ahmed2021}
Ahmed SE, Pawar S, San O, Rasheed A, Iliescu T, Noack BR. 2021  On closures for reduced order models—A spectrum of first-principle to machine-learned avenues. {\em Physics of Fluids} \textbf{33}.
(\href{http://dx.doi.org/10.1063/5.0061577}{10.1063/5.0061577})

\bibitem{Bernd2011a}
Bernd~R. N, Marek M, Gilead T, editors. 2011 {\em Reduced-Order Modelling for Flow Control}.
Berlin, Heidelberg: Springer Vienna.
(\href{http://dx.doi.org/10.1007/978-3-7091-0758-4}{10.1007/978-3-7091-0758-4})

\bibitem{Girfoglio2021}
Girfoglio M, Quaini A, Rozza G. 2021  A POD-Galerkin reduced order model for a LES filtering approach. {\em Journal of Computational Physics} \textbf{436}, 110260.
(\href{http://dx.doi.org/10.1016/j.jcp.2021.110260}{10.1016/j.jcp.2021.110260})

\bibitem{hannachi2007empirical}
Hannachi A, Jolliffe IT, Stephenson DB. 2007  Empirical orthogonal functions and related techniques in atmospheric science: A review. {\em International Journal of Climatology: A Journal of the Royal Meteorological Society} \textbf{27}, 1119--1152.
(\href{http://dx.doi.org/10.1002/joc.1499}{10.1002/joc.1499})

\bibitem{Monahan2009}
Monahan AH, Fyfe JC, Ambaum MHP, Stephenson DB, North GR. 2009  Empirical Orthogonal Functions: The Medium is the Message. {\em Journal of Climate} \textbf{22}, 6501--6514.
(\href{http://dx.doi.org/10.1175/2009JCLI3062.1}{10.1175/2009JCLI3062.1})

\bibitem{Shen1998}
Shen SS, Smith TM, Ropelewski CF, Livezey RE. 1998  An Optimal Regional Averaging Method with Error Estimates and a Test Using Tropical Pacific SST Data. {\em Journal of Climate} \textbf{11}, 2340--2350.
(\href{http://dx.doi.org/10.1175/1520-0442(1998)011<2340:aoramw>2.0.co;2}{10.1175/1520-0442(1998)011<2340:aoramw>2.0.co;2})

\bibitem{Errors_EOF}
North GR, Bell TL, Cahalan RF, Moeng FJ. 1982  Sampling Errors in the Estimation of Empirical Orthogonal Functions. {\em Monthly Weather Review} \textbf{110}, 699 -- 706.
(\href{http://dx.doi.org/10.1175/1520-0493(1982)110<0699:SEITEO>2.0.CO;2}{10.1175/1520-0493(1982)110<0699:SEITEO>2.0.CO;2})

\bibitem{Everson1995}
Everson R, Sirovich L. 1995  Karhunen–Loève procedure for gappy data. {\em Journal of the Optical Society of America A} \textbf{12}, 1657.
(\href{http://dx.doi.org/10.1364/josaa.12.001657}{10.1364/josaa.12.001657})

\bibitem{EOF_Sparse}
Beckers JM, Rixen M. 2003  EOF calculations and data filling from incomplete oceanographic datasets. {\em Journal of Atmospheric and oceanic technology} \textbf{20}, 1839--1856.
(\href{http://dx.doi.org/10.1175/1520-0426(2003)020<1839:ECADFF>2.0.CO;2}{10.1175/1520-0426(2003)020<1839:ECADFF>2.0.CO;2})

\bibitem{Willcox2004}
Willcox K.
In {\em Unsteady Flow Sensing and Estimation via the Gappy Proper Orthogonal Decomposition},.
(\href{http://dx.doi.org/10.2514/6.2004-2415}{10.2514/6.2004-2415})

\bibitem{Tseringxiao2019}
Tsering-xiao B, Xu Q. 2019  Gappy POD-based reconstruction of the temperature field in Tibet. {\em Theoretical and Applied Climatology} \textbf{138}, 1179--1188.
(\href{http://dx.doi.org/10.1007/s00704-019-02898-6}{10.1007/s00704-019-02898-6})

\bibitem{Alvera‐Azcarate2007}
Alvera‐Azcárate A, Barth A, Beckers J, Weisberg RH. 2007  Multivariate reconstruction of missing data in sea surface temperature, chlorophyll, and wind satellite fields. {\em Journal of Geophysical Research: Oceans} \textbf{112}.
(\href{http://dx.doi.org/10.1029/2006jc003660}{10.1029/2006jc003660})

\bibitem{Tipping1999}
Tipping ME, Bishop CM. 1999  Probabilistic Principal Component Analysis. {\em Journal of the Royal Statistical Society Series B: Statistical Methodology} \textbf{61}, 611--622.
(\href{http://dx.doi.org/10.1111/1467-9868.00196}{10.1111/1467-9868.00196})

\bibitem{goodin1979comparison}
Goodin WR, McRa GJ, Seinfeld JH. 1979  A comparison of interpolation methods for sparse data: Application to wind and concentration fields. {\em Journal of Applied Meteorology and Climatology} \textbf{18}, 761--771.
(\href{http://dx.doi.org/10.1175/1520-0450(1979)018<0761:ACOIMF>2.0.CO;2}{10.1175/1520-0450(1979)018<0761:ACOIMF>2.0.CO;2})

\bibitem{Miro2017}
Miró JJ, Caselles V, Estrela MJ. 2017  Multiple imputation of rainfall missing data in the Iberian Mediterranean context. {\em Atmospheric Research} \textbf{197}, 313--330.
(\href{http://dx.doi.org/10.1016/j.atmosres.2017.07.016}{10.1016/j.atmosres.2017.07.016})

\bibitem{Schanz2016}
Schanz D, Gesemann S, Schröder A. 2016  {Shake-The-Box: Lagrangian particle tracking at high particle image densities}. {\em Experiments in Fluids} \textbf{57}.
(\href{http://dx.doi.org/10.1007/s00348-016-2157-1}{10.1007/s00348-016-2157-1})

\bibitem{Tan2020}
Tan S, Salibindla A, Masuk AUM, Ni R. 2020  Introducing {OpenLPT}: new method of removing ghost particles and high‑concentration particle shadow tracking. {\em Experiments in Fluids} \textbf{61}.
(\href{http://dx.doi.org/10.1007/s00348-019-2875-2}{10.1007/s00348-019-2875-2})

\bibitem{Schroeder2023}
Schröder A, Schanz D. 2023  3D Lagrangian Particle Tracking in Fluid Mechanics. {\em Annual Review of Fluid Mechanics} \textbf{55}, 511--540.
(\href{http://dx.doi.org/10.1146/annurev-fluid-031822-041721}{10.1146/annurev-fluid-031822-041721})

\bibitem{wang2017interpolation}
Wang Y, Akeju OV, Zhao T. 2017  Interpolation of spatially varying but sparsely measured geo-data: A comparative study. {\em Engineering Geology} \textbf{231}, 200--217.
(\href{http://dx.doi.org/10.1016/j.enggeo.2017.10.019}{10.1016/j.enggeo.2017.10.019})

\bibitem{sperotto2022meshless}
Sperotto P, Pieraccini S, Mendez MA. 2022  A meshless method to compute pressure fields from image velocimetry. {\em Measurement Science and Technology} \textbf{33}, 094005.
(\href{http://dx.doi.org/10.1088/1361-6501/ac70a9}{10.1088/1361-6501/ac70a9})

\bibitem{ramsay2005principal}
Ramsay J, Silverman B. 2005  Principal components analysis for functional data. {\em Functional data analysis} pp. 147--172.

\bibitem{Ramsay1997}
Ramsay JO, Silverman BW. 1997 {\em Functional Principal Component Analysis}.
Springer.

\bibitem{Wang2016}
Wang JL, Chiou JM, Müller HG. 2016  A Survey of Functional Principal Component Analysis. {\em The American Statistician} \textbf{70}, 127--137.
(\href{http://dx.doi.org/10.1080/00031305.2016.1141704}{10.1080/00031305.2016.1141704})

\bibitem{Hall2006}
Hall P, Horowitz JL. 2006  On Properties of Functional Principal Components Analysis. {\em Journal of the Royal Statistical Society: Series B (Statistical Methodology)} \textbf{68}, 109--126.
(\href{http://dx.doi.org/10.1111/j.1467-9868.2005.00535.x}{10.1111/j.1467-9868.2005.00535.x})

\bibitem{Tripathy2021}
Tripathy B, Anveshrithaa S, Ghela S. 2021  Dual PCA. In {\em Unsupervised Learning Approaches for Dimensionality Reduction and Data Visualization} ,  pp. 19--22. CRC Press.
(\href{http://dx.doi.org/10.1201/9781003190554-3}{10.1201/9781003190554-3})

\bibitem{Ghojogh2022}
Ghojogh B, Crowley M. 2022  Unsupervised and Supervised Principal Component Analysis: Tutorial. {\em arXiv preprint arXiv:1906.03148}.
(\href{http://dx.doi.org/10.48550/arXiv.1906.03148}{10.48550/arXiv.1906.03148})

\bibitem{buhmann2000radial}
Buhmann MD. 2000  Radial basis functions. {\em Acta numerica} \textbf{9}, 1--38.
(\href{http://dx.doi.org/10.1017/S0962492900000015}{10.1017/S0962492900000015})

\bibitem{tirelli2023simple}
Tirelli I, Ianiro A, Discetti S. 2023  A simple trick to improve the accuracy of {PIV/PTV} data. {\em Experimental Thermal and Fluid Science} \textbf{145}, 110872.
(\href{http://dx.doi.org/10.1016/j.expthermflusci.2023.110872}{10.1016/j.expthermflusci.2023.110872})

\bibitem{Numerics}
Griffiths DV, Smith I. 2006 {\em Numerical Methods for Engineers}.
Boca Raton, Florida: Routledge, Taylor \& Francis.
(\href{http://dx.doi.org/10.1201/9781420010244}{10.1201/9781420010244})

\bibitem{Scott}
Dawson S. 2023 p. 117–132.
In {\em The Proper Orthogonal Decomposition}, p. 117–132. Cambridge University Press.
(\href{http://dx.doi.org/10.1017/9781108896214.011}{10.1017/9781108896214.011})

\bibitem{ralston2001first}
Ralston A, Rabinowitz P. 2001 {\em A first course in numerical analysis}.
Courier Corporation.

\bibitem{Liu2005}
Liu G, Gu Y. 2005 {\em An Introduction to Meshfree Methods and Their Programming}.
Springer Netherlands.
(\href{http://dx.doi.org/10.1007/1-4020-3468-7}{10.1007/1-4020-3468-7})

\bibitem{Zhang2024}
Zhang Z, Li M, Wang X. 2024  Review of Collocation Methods and Applications in Solving Science and Engineering Problems. {\em Journal of Computational Physics} \textbf{456}, 110937.
(\href{http://dx.doi.org/10.1016/j.jcp.2023.110937}{10.1016/j.jcp.2023.110937})

\bibitem{deng2020low}
Deng N, Noack BR, Morzy{\'n}ski M, Pastur LR. 2020  Low-order model for successive bifurcations of the fluidic pinball. {\em Journal of fluid mechanics} \textbf{884}, A37.
(\href{http://dx.doi.org/10.1017/jfm.2019.959}{10.1017/jfm.2019.959})

\bibitem{Konread}
Konrad H, Panegrossi G, Bagaglini L, Sanò P, Sikorski T, Cattani E, Schröder M, Mikalsen A, Hollmann R. 2022  Precipitation monthly and daily gridded data from 2000 to 2017 derived from satellite microwave observations. {\em Copernicus Climate Change Service (C3S) Climate Data Store (CDS)}.
(\href{http://dx.doi.org/10.24381/cds.ada9c583}{10.24381/cds.ada9c583})

\end{thebibliography}

\end{document}